\documentclass[prl,twocolumn,aps, showpacs]{revtex4}

\usepackage{epsf,ifthen,epsfig}

\def\ni{\noindent}
\def\simge{%
    \mathrel{\rlap{\raise 0.511ex
        \hbox{$>$}}{\lower 0.511ex \hbox{$\sim$}}}}
\def\simle{%
    \mathrel{\rlap{\raise 0.511ex
        \hbox{$<$}}{\lower 0.511ex \hbox{$\sim$}}}}

\begin{document}

%\title{THE ULTIMATE ENERGY DENSITY OF COLD MATTER}
\title{The Ultimate Energy Density of Observable Cold Matter} 
\author{James M. Lattimer and Madappa Prakash }
\affiliation{Department of Physics and Astronomy, Stony Brook University, 
Stony Brook, NY 11794-3800}

%\date{\today}

\begin{abstract}
We demonstrate that the largest measured mass of a neutron star
establishes an upper bound to the energy density of 
observable cold matter.  An equation of state-independent
expression satisfied by both normal neutron stars and self-bound quark
matter stars is derived for the largest energy density inside
stars as a function their masses.   The largest observed mass sets the
lowest upper limit to the density.  Implications from existing and
future neutron star mass measurements are discussed. \\
\end{abstract}


\vspace{0.4cm}

\maketitle

%\begin{narrowtext}
\newpage

%\section{INTRODUCTION}
The number of neutron stars with measured masses has grown in recent years
\cite{Thorsett99,Stairs04}. The most accurately measured masses are from
timing observations of radio binary pulsars \cite{Manchester04} and, until
recently, were consistent with neutron star masses in the range 1.26 to 1.45
M$_\odot$ \cite{Thorsett99}. Recent data on 
binaries containing
pulsars and white dwarfs, however, indicate a larger range of
masses \cite{Stairs04}. An example is the binary containing PSR J0751-1807,
with $2.2\pm0.2$ M$_\odot$ with 1$\sigma$ errors \cite{Nice03a}. Data from
x-ray binaries \cite{Thorsett99} also suggest a wide range in masses, but
are subject to greater theoretical and observational uncertainties. As
neutron stars are expected to contain the densest cold matter outside black
holes, the maximum neutron mass mass and the corresponding maximum energy
density are of great interest.

We demonstrate here that a precisely measured neutron star mass sets
an upper limit to the mass density, or, equivalently, the energy
density, inside the star. The larger the measured mass, the smaller
the density limit.  A sufficiently large mass could 
%eliminate 
delimit classes
of possible equations of state (EOS's).  A limit for this
maximum density is proferred 
utilizing an analytic solution of
Einstein's equations.  This limit is checked by comparing
numerical results for a variety of EOS's of both normal and self-bound
stars.  Recent neutron star mass measurements are summarized and
inferences drawn.

From the general relativistic structure equations
\cite{Tolman39,Oppenheimer39}, the maximum compactness of a star is
set by the limit $R>(9/4)GM/c^2$ \cite{Buchdahl5966}, where
$R$ and $M$ are the stellar radius and mass, respectively.  With the
additional requirements that (i) nowhere in the star is the speed of
sound $c_s$ greater than the speed of light $c$, (ii) $c_s$ is
everywhere real, and (iii) the EOS matches smoothly to calculable low
density EOS's near the nuclear saturation density
$\rho_s\simeq2.6\times10^{14}$ g cm$^{-3}$, Ref. \cite{Glendenning92}
showed that the compactness limit is increased to
\begin{equation}
R\simge2.94GM/c^2\,.
\label{eq:caus}
\end{equation}  
This result improves the limit $R\simge3.05GM/c^2$ established
\cite{Lattimer90} using the prescription $c_s=c$ above a fiducial
energy density $\rho_f$ \cite{Rhoades74}.  The maximum mass inferred
from this prescription is proportional to $\rho_f^{-1/2}$, but the
compactness limit is independent of $\rho_f$ for $\rho_f<<\rho_c$,
where $\rho_c$ is the central density of the star \cite{Lattimer90}.

The central mass density of a star must be greater than the
average density $\rho_*=3M/(4\pi R^3)$, the value for 
a uniform density star with the same mass and radius.  Combining the
compactness limit, Eq. (\ref{eq:caus}), with the constant density relation
$\rho_*=\rho_c$, yields
\begin{eqnarray}
\rho_{c,*}
%&=&{3\over4\pi}\left({c^2\over2.94G}\right)^3{1\over M^2}\cr
\simeq 5.80\times10^{15}~\left({{\rm M}_\odot/ M}\right)^2 
{\rm~g~cm}^{-3}\,.
\label{eq:inc}
\end{eqnarray}
This is a plausible approximate lower limit to the central density $\rho_c$
for a star of a given mass, but it is not an absolute lower limit. (This lower
limit cannot be made firm as causality has been imposed on a uniform density
fluid in which transmission of signals is instantaneous.) A firm lower limit
can be established, however, if an upper limit to $R$ exists. One
observational limitation originates from the most rapidly spinning pulsar,
PSR B1937+21 \cite{Ashworth83}, 
which has a frequency $\nu=641$ Hz. This leads to a lower limit to $M/R^3$
%\cite{LHZ04} 
\cite{Lattimer04} and a lower limit 
\begin{equation}\label{eqn:rot}
\rho_{c,rot}\simeq 1.79\times10^{14} (\nu/641{\rm~Hz})^2
{\rm~g~cm}^{-3}\,,
\end{equation}
which is, however, not very restrictive.  A far more stringent limit
could be achieved from a redshift observed from a neutron star. The
largest observed redshift $z_{obs}$ sets a lower limit to $M/R$, implying
\begin{equation}\label{eqn:z}
\rho_{c,z} > {3\over4\pi M^2}\left({c^2z_{obs}(2+z_{obs})
\over2G(1+z_{obs})^2}\right)^3  \,. 
\end{equation}
Recently, $z_{obs}= 0.35$ was reported \cite{Cottam02} for the
x-ray bursting source XTE J1814-338.  With this value,
\begin{equation}\label{eqn:zobs} 
\rho_{c,z} > 1.69~\times 10^{15}~ ({\rm M}_\odot/M)^2
{\rm~g~cm}^{-3}\,.
\label{eq:zcon}  
\end{equation}

The central question is, how much greater can $\rho_c$ be compared to
any of the above expressions for physically motivated EOS's? If this
question can be answered, an upper limit to the density inside a star
of a given mass can found.  An important consequence of the existence
of an upper limit is that the largest measured neutron star mass would
set an upper limit to the density of cold matter. (In a dynamical
environment, such as the gravitational collapse of a stellar core
to a black hole or a high energy heavy ion collision, matter
becomes hot and may achieve higher densities.) An additional
consequence is that one could infer whether or not non-nucleonic
degrees of freedom, such as hyperons, Bose condensates or quarks,
which generally reduce the maximum mass, can exist in the cores of
neutron stars.

%\subsection{Analytic Solutions}

Some insights can be gained by comparing analytical solutions to
Einstein's equations with numerical solutions employing model
EOS's. The known analytic solutions fall into two classes: (i) the class that
describes ``normal'' neutron stars for which $\rho_c$ vanishes at the
surface where the pressure vanishes, and (ii) the class that describes
``self-bound'' stars for which $\rho_c$ is finite at the surface.  In
the first class, there are only three known analytic solutions: the
Tolman VII solution \cite{Tolman39}, Buchdahl's solution
\cite{Buchdahl67}, and the Nariai IV solution
\cite{Nariai5051}. In the second class, an infinite number of
analytic solutions exist, but the useful ones are variants of the
Tolman IV and VII solutions \cite{Tolman39,Lake03}, as well as
the uniform density case \cite{Schwarzschild16}.

All known analytic solutions are scale-free; they depend
parametrically on the compactness ratio $\beta=GM/Rc^2$. However, by
coupling these solutions with Eq. (\ref{eq:caus}), {\it i.e.,} by
setting the compactness $\beta=\beta_c\equiv1/2.94$, one can obtain
relations between $\rho_c$ and $M$ analogous to that of a uniform
density fluid, Eq. (\ref{eq:inc}).  We thus need to relate the central
density to $\beta$ and $M$ for these solutions.  These are summarized
below: \\

\ni{\em 1. Tolman VII:} 
This solution stems from the ansatz \cite{Tolman39} 
\begin{equation}
\rho=\rho_c\left[1-(1-w)(r/R)^2\right]\,,
\end{equation}
where the parameter $w=\rho_{s}/\rho_c$ is the ratio of the energy
densities at the surface and the center, which can vary between 0 and 1.
(The case $w=1$ represents the uniform density fluid.)  This leads
to
\begin{eqnarray}
\rho_{c,VII}={15\over4\pi(2+3w)}
\left({c^2\beta_c\over G}\right)^3{1\over M^2}\cr
\simeq{1.45 \times 10^{16}\over(1+1.5w)}
\left({{\rm M}_\odot\over M}\right)^2 {\rm~g~cm}^{-3}\,.
\label{TVII}
\end{eqnarray}
This solution is valid, in the case $w=0$ (the normal neutron star
case), for $\beta<0.3862\simeq1/2.59$.  For positive $w$ (the
self-bound case), the solution is valid for larger values of $\beta$.
Thus, this solution is useful for the case $\beta=\beta_c$.  
For $w>0$, the central density decreases for a given mass relative to
the normal neutron star case. \\

\ni{\em 2. Buchdahl:} This solution uses the EOS ansatz 
$\rho c^2=12\sqrt{p_*P}-5P$, where $p_*$ is a constant.  In this case
\cite{Buchdahl67}
\begin{eqnarray}
\rho_{c,Buch}&=&{\pi(2-5\beta_c)(1-\beta_c)^2\over8(1-2\beta_c)}
\left({c^2\beta_c\over G}\right)^3{1\over M^2}\cr
&\simeq&3.89\times10^{15}~\left({{\rm M}_\odot/ M}\right)^2 {\rm~g~cm}^{-3}.
\label{Buchdahl}
\end{eqnarray}
However, because Buchdahl's solution is invalid when
$\beta\ge1/5$, the value for which the central sound speed becomes
infinite, it cannot be used for the case $\beta=\beta_c$. \\

\ni{\em 3. Nariai IV:} This solution is characterized by 
\cite{Nariai5051}
\begin{eqnarray}
\rho_{c,NIV}&=&{3\over8\pi}\left[(\alpha-1)\cos\sqrt{3\beta_c}+
{(6-\alpha)\over\sqrt{3\beta_c}}\sin\sqrt{3\beta_c}\right]\cr
&&\times\left({c^2\beta_c\over G}\right)^3{1\over M^2}\cr
&\simeq&9.88\times10^{15}~\left({{\rm M}_\odot/ M}\right)^2 
{\rm~g~cm}^{-3}.
\label{NIV}
\end{eqnarray}
Here,
\begin{eqnarray}\label{ndef}
\alpha={12+\beta_c\over2+\beta_c+2\sqrt{1-2\beta_c}}\,.
\end{eqnarray}
This solution is valid for $\beta<0.4126$, the value for which the central 
pressure and sound speed become infinite.  This solution can also be
generalized to include self-bound stars, and as for the Tolman VII
case, the central density for a given mass decreases from that of
Eq. (\ref{NIV}) as the ratio $w$ is increased from 0. \\

\ni{\em 4. Tolman IV (generalized):}
Lake \cite{Lake03} showed that the ansatz for the metric function
\begin{equation}\label{tolans}
e^{\nu(r)}={[N-\beta(2N+1)+\beta (r/R)^2]^N\over N^N(1-2\beta)^{N-1}}\,,
\end{equation}
where $N$ is a positive integer, produces an infinite family of
analytic solutions of the self-bound type. Four of these were
previously known ($N=1, 3, 4$ and 5). The case $N=1$ cannot properly
be applied to our problem as this solution is finite only for
$\beta<1/3$. The most relevant case is for $N=2$, for
which $c_s \approx \sqrt{1/3}$ throughout most of the star, similar
to the behavior of strange quark matter. For this case,
\begin{eqnarray} 
\rho_{c,TIV}&=&{3\over4\pi}
\left({2-2\beta_c\over2-5\beta_c}\right)^{2/3} \left({c^2\beta_c\over
G}\right)^3{1\over M^2}\cr
&\simeq&1.56\times10^{16}~\left({{\rm M}_\odot/
M}\right)^2 {\rm~g~cm}^{-3}\,.
\label{TIV} 
\end{eqnarray}
This solution is valid for $\beta<2/5$.  The ratio of the surface to the
central densities for the case $N=2$ is
\begin{eqnarray}
w={6-10\beta\over3}\left({2-5\beta\over2-2\beta}\right)^{2/3}
\end{eqnarray}
which is approximately 0.32 for $\beta=\beta_c$.
With increasing $N$, the central density for a
given mass decreases from that of Eq. (\ref{TIV}), and in the limit
$N>>1$, $\rho_{c,TIV}(M)$ approaches the uniform density result.

\begin{figure}[hbt!]
\begin{center}
\epsfig{file=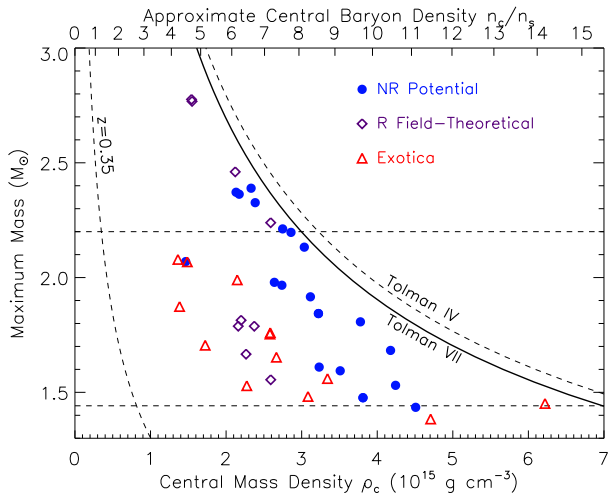, angle=90, height=2.6in}
\end{center}
\vskip -10pt
\caption{The central energy density and mass of maximum mass
configurations. Symbols reflect the nature of the EOS's selected from
Refs. \cite{Lattimer01}. NR are non-relativistic
potential models, R are field- theoretical models, and Exotica refers
to NR or R models in which strong softening occurs, due to the
occurence of hyperons, a Bose condensate, or quark matter.  The
Exotica points include self-bound strange quark matter stars.  For
comparison, the central density -- maximum mass relations for the 
%constant density, 
redshift, 
Tolman VII
($w=0$), and Tolman IV ($N=2$) 
bounds from Eqs.~(\ref{eq:zcon}), (\ref{TVII}), and
(\ref{TIV})
are shown.  The
dashed lines for 1.44 and 2.2 M$_\odot$ serve to guide the eye.}
\vskip -10pt
\label{fig:mmax}
\end{figure}

To investigate the relevance of analytic relations between the central
density and mass, we carried out numerical integrations of the TOV
structure equations for a multitude of EOS's, including potential and
field-theoretical models, and models that contain strong softening due
to the occurence of hyperons, Bose condensates or quark matter,
including the case of self-bound strange quark matter stars. The EOS's
were chosen from Refs.  \cite{Lattimer01}. Fig. \ref{fig:mmax}
displays the maximum masses and the central energy densities of the
maximum mass configurations.  Analytic solutions for the Tolman VII
normal neutron star case ($w=0$), the Tolman IV case for $N=2$, and
the uniform density case are also displayed. The paths of other
analytic solutions are not shown for clarity; they scale up or down
from the analytic results shown (all analytic solutions for fixed
$\beta$ obey the law $\rho_c\propto M^{-2}$).

It is fortuitous but significant that the Tolman VII solution forms
a strict upper limit to the density of a maximum mass star, for each
of the EOS's displayed. We therefore conjecture that the Tolman VII
curve marks the upper limit to the energy density inside a star of the
indicated mass. Since the maximum density achieveable with a given EOS
is the central density of the maximum mass star, a stellar mass
measurement can be directly converted into an upper limit for the
maximum density. Since the measured mass must necessarily be less than
the neutron star maximum mass, this limit also forms an absolute upper
density inside any compact star. In other words, except in transient
situations such as the Big Bang or in relativistic heavy ion
collisions, this curve displays the ultimate density of matter.

%\vskip-5pt
\begin{figure}[hbt!]
%\hskip -.835in
%\includegraphics[scale=.5]{masses.ps}
%
\begin{center}
\hskip -.5in\epsfig{file=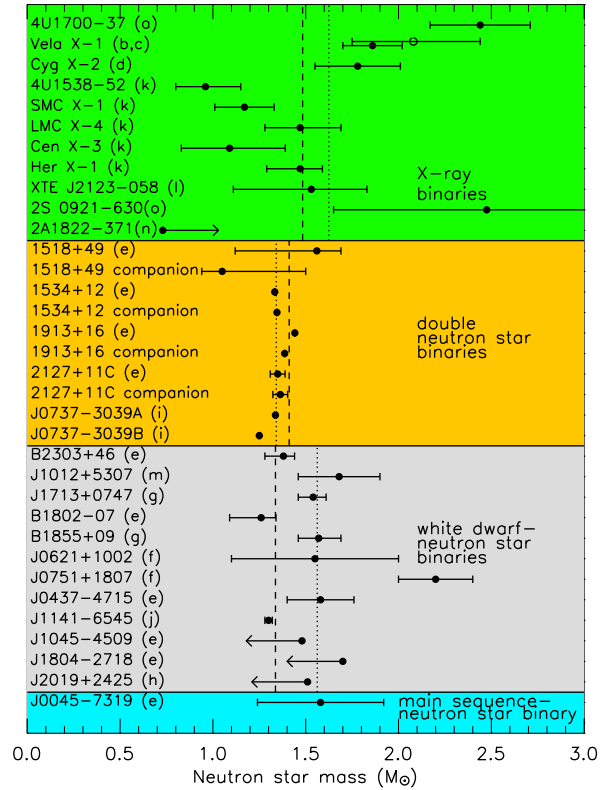, angle=0, width=0.52\textwidth}
\end{center}
\vskip-10pt
\caption{Measured and estimated masses of neutron stars in radio
binary pulsars and in x-ray accreting
binaries.  Sources are listed by letter
(Refs. \cite{Stairs04,Nice03a} and [20-35]).  Error bars are
$1\sigma$.  Vertical dotted lines show average masses of each group
(1.62 M$_\odot$, 1.34 M$_\odot$ and 1.56 M$_\odot$); dashed vertical
lines indicate inverse error weighted average masses (1.48 M$_\odot$,
1.41 M$_\odot$ and 1.34 M$_\odot$).}
\vskip-10pt
\label{fig:masses}
\end{figure}

The above results are given in terms of the central mass or energy
density $\rho$.  However, most models of dense matter
are formulated in terms of the baryon number density $n$.  Therefore,
it is of interest to examine relationships between the central baryon
density and mass for the maximum mass configurations.  A good rule of
thumb for converting $\rho$ to $n$, using $n_s\simeq0.16$ fm$^{-3}$, is
\begin{equation}
\rho/\rho_s \approx 0.9 (n/n_s) 
[1 + 0.11 (n/n_s)^{3/4}] \,. 
\end{equation}
The number density so obtained is indicated on the top scale of
Fig.~\ref{fig:mmax}.  We emphasize that the plotted points are positioned
using $\rho_c$, not $n_c$, in this figure.

The most accurately measured neutron star masses are from timing observations
of radio binary pulsars \cite{Manchester04,Stairs04}. These include
pulsars orbiting another neutron star, a white dwarf or a main-sequence star.

Measured masses are summarized in Fig. \ref{fig:masses} and
are plotted with $1\sigma$ uncertainties. Ordinarily, observations of pulsars
in binaries yield orbital sizes and periods from Doppler phenomenon, from
which the total mass of the binary can be deduced. But the compact nature of
several binary pulsars permits detection of relativistic effects, such as
Shapiro delay or orbit shrinkage due to gravitational radiation reaction,
which constrains the inclination angle and permits measurement of each mass in
the binary. The largest accurately measured mass originates from the binary
pulsar system PSR 1913+16, whose masses are $1.3867\pm0.0002$ and
$1.4414\pm0.0002$ M$_\odot$, respectively \cite{Weisberg04}.

A significant development concerns mass determinations in binaries
with white dwarf companions, which show a broader mass range than
binary pulsars having neutron star companions. Ref. \cite{Bethe98}
suggests that a narrow set of evolutionary circumstances conspire to
form double neutron star binaries, leading to a restricted range of
neutron star masses.  The implication of this restriction for other
binaries remains to be explored.  A few cases of white dwarf binaries
that contain neutron stars considerably larger than the canonical 1.4
M$_\odot$ value have been reported. A striking case is PSR J0751+1807
\cite{Nice03a} in which the estimated mass with $1\sigma$ error
bars is $2.2\pm0.2$ M$_\odot$. For this neutron star, a mass of 1.4
M$_\odot$ is $4\sigma$ away.  If this mass determination holds up
after further observations, the central density constraints become
intriguingly close to the estimated density for the quark-hadron phase
transition.  Raising the limit for the neutron star maximum mass could
also mark the boundaries of  
%eliminate 
other families of EOS's
in which
substantial softening begins around 2 to $3n_s$. This is significant,
since exotica 
generally reduce
the maximum mass appreciably.

The simple mean of the measured neutron star masses in white dwarf-neutron
star binaries exceeds that of the double neutron star binaries by about 0.22
M$_\odot$ (Fig. \ref{fig:masses}). Nevertheless, caution is in order since the
$2\sigma$ errors of most of these systems extend below 1.4 M$_\odot$.
Continued observations are required to reduce these errors.

Masses can also be estimated for
binaries which contain an accreting neutron star emitting x-rays. Some of
these
stars are characterized by relatively large masses but also large
estimated errors (Fig. \ref{fig:masses}). The system Vela X-1
is noteworthy,
because its lower mass limit (1.6 to $1.7$ M$_\odot$) is constrained,
albeit mildly, by geometry \cite{Quaintrell03}. The source 
4U 1700-37 might be a black hole, due to lack of oscillations in
its x-ray spectrum \cite{Clark02}. Another object,
2S 0921-630 \cite{Jonker04}, could either be a high-mass neutron star
or a low-mass black hole. Although not yet demonstrated, it is widely
believed that black holes formed in gravitational collapse have masses
that exceed the neutron star maximum mass. These latter two objects
could play a significant role in determining the neutron star maximum
and the black hole minimum masses.

This work was supported in part by USDOE grant DE-FG02-87ER-40317.

%\end{narrowtext}

\end{document}